\documentclass[aps,prl,twocolumn,groupedaddress]{revtex4}

\usepackage{graphicx}

\begin{document}
Published in \emph{Physical Review Letters}, 103, 215502 (2009)
\title{Self-sustained Levitation of Dust Aggregate Ensembles by Temperature Gradient Induced Overpressures}

\author{Thorben Kelling}
\email[]{thorben.kelling@uni-due.de}
\affiliation{Institut f\"{u}r Planetologie, Universit\"{a}t M\"{u}nster, Wilhelm-Klemm-Strasse 10, 48149 M\"{u}nster, Germany}
\author{Gerhard Wurm}
\affiliation{Fakult\"{a}t f\"{u}r Physik, Universit\"{a}t Duisburg-Essen, Lotharstr. 1, 47048 Duisburg, Germany}


\date{\today}

\begin{abstract}
In laboratory experiments we observe dust aggregates from $100$ $\mu$m to 1 cm in size composed of micrometer 
sized grains
levitating over a hot surface. Depending on the dust sample aggregates start to levitate at a temperature of 400 K. Levitation of dust aggregates is restricted to a pressure range 
between 1--40 mbar. The levitating is caused by a Knudsen compressor effect.
Based on thermal transpiration through the dust aggregates the pressure increases between surface and aggregates. Dust aggregates are typically balanced $\sim$100 $\mu$m over the surface.
On a slightly concave surface individual aggregates are trapped at the center. Ensembles of aggregates 
are confined in a 2D plane. Aggregates are subject to systematic and random translational and 
rotational motion. The levitated aggregates are well suited to study photophoretic or thermophoretic forces on dust aggregates or the mutual interaction between dust aggregates.
\end{abstract}

\maketitle

\section{Introduction}

\begin{figure}
\begin{center}
\includegraphics[width=0.45\textwidth]{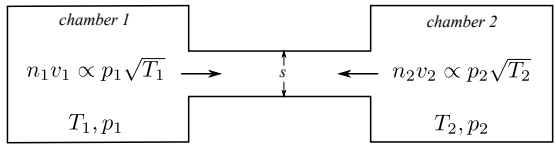}
\end{center}
\caption{Two chambers 1 and 2 on different temperatures $T_1$ and $T_2>T_1$ are connected by a tube with diameter $s$. If $s$ is small or comparable to the mean free path $\lambda$ of the gas molecules ($s\ll\lambda$) thermal creep results in a higher pressure in chamber 2 compared to chamber 1 where $nv$ is the particle flow rate per cross section, $n$ is the gas molecules number density and $v$ is the gas molecules mean velocity.}
\label{knudsenoriginal}
\end{figure}

The properties of nano- and micrometer-sized particles (dust) and aggregates composed of dust 
grains are an active field of investigation in many scientific areas. Collisional behaviour of dust aggregates at low velocities is important in astrophysics, i.e. planet formation \cite{ormel,brauer08}. Optical properties are important for atmospheric science \cite{schmid,munoz}, and thermal properties are important for effects like photo-induced erosion of dust beds \cite{wurm06,wurm08}. Dust aggregate properties are not always easily accessible. One main problem for experimental studies is to provide individual 
or ensembles of free dust aggregates for experiments. So far, i.e., slow collisions ($v_{col}<1$ m/s) between aggregates are usually studied under microgravity as no suitable technique exists for ground based studies \cite{langkowski, blum93}. In this article the levitation of hot dust aggregates in a low pressure gaseous environment is presented. The levitation is based on a Knudsen compressor effect which is depicted in Fig.\ref{knudsenoriginal} \cite{knudsen09}. If the connection with diameter $s$ of two connected gas reservoirs on different temperatures $T_1$ and $T_2>T_1$ is small compared to the mean free path of the gas molecules ($s\ll\lambda$), an overpressure on the warmer side will be established. In equilibrium it is $p_2/p_1=\sqrt{T_2/T_1}$. In non-equilibrium with $s\simeq \lambda$ the overpressure is \cite{vargo99,muntz02}

\begin{equation}
\Delta p = p_{avg}\frac{Q_T}{Q_P}\frac{\Delta T}{T_{avg}}.
\label{overpressure}
\end{equation} 

Here $p_{avg}$ and 
$T_{avg}$ are the average pressure and temperature, $Q_T/Q_p$ is the ratio between the transpiration (creep) and back flow 
of the gas and $\Delta T = T_2-T_1$ is the temperature difference between the two reservoirs.  

\section{Experiments}

In experiments we placed dust aggregates onto a heater and adjusted the ambient pressure to about 10 mbar. 
The dust is then heated to more than 400 K when dust aggregates start to levitate. 
On a slightly concave heater individual aggregates are confined to the vicinity of the center. 
An ensemble of aggregates is also well confined. 
Dust aggregates interact with each other at low collision velocities.
Typical parameters of the used dust samples are given in 
Table \ref{dustprop}.

\begin{table}
\caption{Dust Properties and typical levitation pressure range 
where most aggregates of a sample are levitated.
Heater temperature is $\sim$800K.}
\begin{tabular}{lccl}
\hline\hline
Dust&Grain Size&Density&Pressure\\
&$\rm \mu m$&$\rm g/cm^3$&mbar\\
\hline

$SiO_2$\\

\quad quartz&$0.1 - 10$&2.6&$3-10$\\
\quad spherical& 1.2&2.0&$7-10^*$\\
\quad Cabosil M5&$<1$&$2.2$&$1-40$\\

$TiO_2$&$< 1$&3.9&$3-30^*$\\
$SiC$&$5 - 15$&3.2&$4-35$\\

$basalt$& $< 100$ &$3.3$&$2-20$\\
$graphite$& $< 20$&2.2&$2-50$\\
$iron$&$6-9$&7.9&no lev.\\
\\
$glass$\\
\quad single sphere&1000&2.5&no lev.\\
\quad single sphere&$40-70$&2.5&no lev.\\
\quad aggregate&$40-70$&2.5&$2-7^*$\\

\hline\hline
\multicolumn{4}{l}{$^*$some movement but no continuous levitation.}
\end{tabular}
\label{dustprop}
\end{table}

\begin{figure}
\begin{center}
\includegraphics[width=0.45\textwidth]{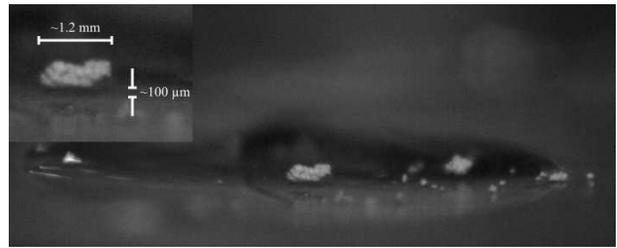}
\end{center}
\caption{Levitation of a 1.2 mm SiO$_2$ aggregate at 3 mbar and $\sim$800 K heater temperature. The reflection of the aggregate on the concave heater surface can be seen distinct from the aggregate. The levitation height is about $\sim$100 $\mu$m.
Other aggregates visible are not levitated.}
\label{levitation}
\end{figure}

With the exception of iron, all dust samples were successfully trapped levitating at the center of the heater. Individual glass spheres of 40--70 $\mu$m were not levitated but larger ($> 100$ $\mu$m) ensembles of such spheres in close contact -- while not quite levitated -- were moving in agreement to the model discussed below. The confinement of the aggregates is robust and stable. The aggregates can be trapped levitating at least 1 minute, which is the longest we did run an experiment.
An example of a levitated SiO$_2$ aggregate is shown in Fig.\ref{levitation}. The typical levitation height is on the order of $\sim$100 $\mu$m, varying slightly for different temperatures and pressures. 
This height is not very specific for a chosen dust sample and its aggregate size, i.e. the heights of aggregates with lateral extensions between $100$ $\mu$m and 1 cm are comparable. Individual aggregates are trapped near the center of the slightly concave heater. However, aggregate rotation around a vertical axis frequently occurs. Typical rotation frequencies are $\nu<10$ Hz. Also, periodic movements, e.g. elliptical movement around or oscillations through the center of the heater are possible. Smaller aggregates ($< 100$ $\mu$m) are sometimes lifted higher or even ejected from a mm-sized dust aggregate. We attribute this to photophoretic forces acting on the total but small aggregate. In contrast we attribute the hovering of larger aggregates to a pressure difference generated in analogy to a Knudsen compressor by thermal transpiration (see model below) \cite{knudsen09}. 
For all dust samples and aggregates, the maximal levitation height and hence the strongest Knudsen compressor effect is found at pressures between 1 and 10 mbar.

\begin{figure}
\begin{center}
\includegraphics[width=0.45\textwidth]{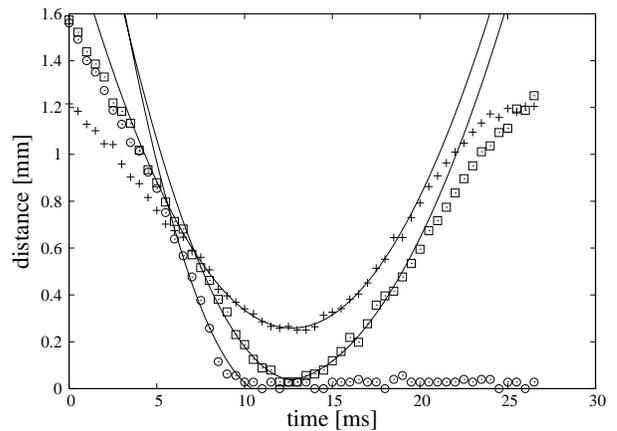}
\end{center}
\caption{Distance between the closest points of interacting aggregates over time for three events. The solid lines represent parabolic fits at approach and reproach. Crosses: repulsive approach of aggregates of equivalent radii  1.3 mm, 0.5 mm; squares: rebound (1.3 mm, 0.6 mm); circles: sticking (0.6 mm, 0.6 mm)}
\label{collisions}
\end{figure}

If the heater is on while several dust aggregates are present, an ensemble of aggregates is levitated and 
interactions between the aggregates occur. The aggregates move partially random translational as well
as rotational within the concave heater. This random motion leads to frequent collisions of aggregates. Collision velocities are typically $v_{col}\ll 1$ m/s. We observed that collisions either lead to growth, i.e. two or more aggregates are merged or aggregates bounce off each other (Fig.\ref{collisions}). Collisions are usually not purely ballistic but repulsive forces are often decelerating aggregates upon approach or even preventing collisions (Fig.\ref{collisions}). Especially for larger aggregates ($>1$ mm), tensions within the aggregate can lead to spontaneous fragmentation.

\section{Model}

\begin{figure}
\begin{center}
\includegraphics[width=0.45\textwidth]{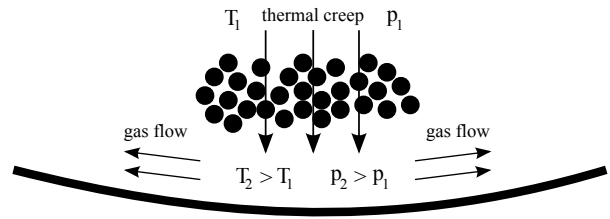}
\end{center}
\caption{A dust aggregate is placed on a slightly concave heater placed in a vacuum chamber which is evacuated to pressures of 10 mbar. The heater is set to $T_2>400$ K. 
Thermal radiation cools the surface of the aggregate to $T_1$. The aggregate is levitated by a pressure difference $\Delta p$ established through thermal creep (Knudsen compressor effect).}
\label{balances}
\end{figure}

A model for the levitation mechanism is shown in Fig.\ref{balances}.
The dust aggregates in the experiments are composed of individual dust grains. Between these grains pores of 
about the size of the grains exist (see Tab.\ref{dustprop}). These pore sizes are on the order of the mean free path $\lambda$ of the gas molecules at the pressures for which levitation is observed (at 10 mbar it is $\lambda\simeq 7$ $\mu$m). If the top and bottom of these pores
is at different temperatures thermal creep leads to a gas flow to the warmer side, in this case to the bottom
of the aggregates \cite{knudsen09,vargo99}. We assume, that the dust aggregates in our experiments act 
as a collection of micro channels separating the gas below the aggregate from the gas above the aggregate. 
If the temperature difference over the aggregate is large enough, the pressure support according to eq.\ref{overpressure} is larger than gravity and the aggregate is lifted. 
The pressure will decrease below the aggregate as gas moves to the sides. The aggregate therefore only rises
until an equilibrium of the inflowing gas through the aggregate and the outflowing gas to the 
open sides below the aggregate is reached. Aggregates are typically levitated at a height which is small compared 
to the lateral extension of the heater (levitation height 100 $\mu$m $\ll$ 1.5 cm bowl diameter). Therefore, the 
bottom temperature is determined by the thermal irradiation from below.
Assuming perfect absorption, the bottom temperature equals the temperature of the heater $T_2$. 
We further assume an aggregate thermal conductivity of $\kappa = 0.1$ W/mK \cite{presley97}. 
Radiation transport through the aggregate can be treated as an effective thermal conductivity \cite{bauer93}. It is $\kappa_R = 4 \sigma \epsilon n^2
T^3 s$ where $\sigma=5.67\times 10^{-8}$ W/(m$^2$K$^4$) is the Stefan-Boltzmann constant, $n$ is the refractive index of the gas, $\epsilon$ is the
emissivity and $s$ is the pore size.
For $\epsilon=1$, $n=1$, $T=800$ K, and $s=10$ $\mu$m we get 
$\kappa_R=0.001$ W/mK. This is two orders of magnitude lower than
the assumed value for the aggregate thermal conductivity. Therefore, radiation transport is not significant in this context.

At the pressure of 10 mbar the top surface of an aggregate cools to $T_1$ by radiation until radiative loss is accounted for by conduction through the aggregate of thickness $d$ and it is $\sigma T_1^4 =\kappa (T_2 - T_1)/d$. For $d = 100$ $\mu$m and $T_2 = 800$ K it is $T_1 = 780$ K. The overpressure $\Delta p$ induced by the Knudsen compressor effect can lift the aggregate if the pressure difference $\Delta p$ acting on the aggregates surface $A=a^2$ equals gravity $F_G = a^2d \rho_d fg$. Here, $A$ is the surface of the aggregate with lateral extension $a$ facing the overpressure, $\rho_d = 2600$ kg/m$^3$ is the density of the bulk material (SiO$_2$ assumed for the calculations), $g=9.81$ m/s$^2$ is the gravitational acceleration, $f=0.3$ is the filling factor of the aggregate. The filling factor might vary significantly for individual aggregates. Considering an aggregate thickness of $d=100$ $\mu$m, a minimum pressure difference $\Delta p_{min} \simeq 1$ Pa is sufficient to compensate gravity. With $Q_T/Q_P=0.2$, an average pressure of $p_{avg} = 10^{3}$ Pa, a temperature difference $\Delta T = 20$ K and an average temperature $T_{avg} = 790$ K, the overpressure $\Delta p$ is about 5 Pa. As the effect is based on pressure, the lift is independent of the lateral extent of the aggregates.
The pressure difference calculated is a factor $\Delta p / \Delta p_{min}\simeq 5$ more than needed and aggregates
are lifted even if conditions are less perfect than assumed or a partial pressure feedback from below the aggregate to the top exists. 
With increasing thickness the temperature difference increases as well which increases the pressure difference.
Therefore, in certain limits, lift is also possible for different aggregate thickness's as it was observed in the experiments. A dust aggregate rises up to a height at which no further overpressure can be established because the gas below the aggregate is not enclosed but can escape to the sides. Furthermore it is observed, that a repulsive force typically decelerates aggregates while approaching each other (Fig.\ref{collisions}). This is in agreement to an outflow of gas at the sides of a levitating aggregate. As two aggregates approach each other this gas flow leads to a deceleration of the approaching aggregates. In addition, the Knudsen compressor model is consistent with individual solid glass spheres not being lifted in contrast to aggregates of these spheres as the latter have pores between them.
Iron powder could not be levitated. However, it is denser and has a higher thermal conductivity than the quartz used for the calculations above. It is consistent with the model
that it is beyond the capability of the trap to lift iron aggregates.

The Knudsen compressor balance is based on the thermal radiation of the heater. We give an estimate of photophoretic and thermophoretic forces induced by this radiation. For a spherical aggregate of radius $a$ the photophoretic force is \cite{roha95}

\begin{eqnarray}
F_{ph}&=&2F_{max}\left(\frac{p_{max}}{p}+\frac{p}{p_{max}}\right)^{-1}\\
p_{max}&=&\frac{\eta}{a}\sqrt{\frac{12RT_{avg}}{M}}\\
F_{max}&=&\frac{\pi\eta a^2 \sigma T_2^4}{2\kappa}\sqrt{\frac{R}{3T_{avg}M}}.
\end{eqnarray}

It is $R=8.3$ J/(mol$\,$K), $a$ the particle radius, $\eta=1.8\times10^{-5}$ kg/(ms) the gas viscosity, $T_{avg}=790$ K the gas temperature and $M = 29\times 10^{-3}$ kg/mol the molar mass of the air. For small particles ($a=10$ $\mu$m) it is $F_{ph}/F_G\simeq 4$ while for mm-particles it is $F_{ph}/F_G \simeq 10^{-3}$. Photophoresis should therefore be capable of lifting small aggregates ($< 100$ $\mu$m). Indeed we observed the ejection of smaller aggregates in the experiments, consistent with photophoretic lift. A similar photo-induced ejection mechanism was discussed by \cite{wurm06, wurm08}. We also observed small aggregates levitating at heights larger than $100$ $\mu$m, i.e. at mm-heights. This is consistent with thermal radiation and photophoresis only decreasing significantly in strength further away from the heater. The thermophoretic force for mm-aggregates in the continuum regime is $F_{th} = (f_{th}a^2\kappa_g (dT/dx))/(\sqrt{2k_B T_{avg}/m_g})$ \cite{zheng02}, with $f_{th}\simeq 0.024$ as dimensionless thermophoretic force, $\kappa_g = 0.01$ W/mK as the thermal conductivity of the gas, $dT/dx=2\times 10^5$ K/m as the temperature gradient over the aggregate, $k_B = 1.38\times 10^{-23}$ J/K as Boltzmann constant and $m_g=4.8\times10^{-26}$ kg as the air molecular mass one gets a ratio of $F_{th}/F_G\simeq 10^{-3}$. The levitation of mm aggregates, as observed in the experiments, can therefore not be explained by photophoretic or thermophoertic forces.
In any case thermophoresis or photophoresis would typically not balance aggregates as close as $100$ $\mu$m over the surface as
it has no regulating gas flow.

\section{Applications and conclusion}
It was demonstrated that aggregates with lateral extension of 100 $\mu$m to 1 cm can be levitated over a hot surface through thermal transpiration (Knudsen compressor). The confining and repulsive forces are
relatively weak, readily allowing collisions at low velocities to occur. This allows the study of a quasi 2D free sample of uncharged aggregates at low pressure.  
By placing several dust aggregates onto the slightly concave heater, low velocity collisions of hot dust aggregates can be studied in detail. Dust growth through collisions of small aggregates is the first stage of planet formation in protoplanetary disks \cite{blum08}. Eventually, these aggregates get compacted in collisions to mm-size aggregates similar to the aggregates studied in this paper. Planet formation models assume that these aggregates continue to collide, stick and grow. However, mm-size aggregates so far were only found to rebound below 1 m/s \cite{blum08}. Collision velocities accessible in previous work for detailed studies were restricted to velocities larger than 0.15 m/s in spite of microgravity conditions \cite{blum93}. For detailed
models of planet formation it is important to know if collisions lead to sticking or rebound and at
what probabilities.
With the levitated ensembles of dust aggregates we are able to
study multitudes  of collisions of mm-aggregates with velocities of $v_{col}\ll 1$ m/s in the laboratory. 
The first ensemble experiments show that if collisions appear, rebound is by far the most probable outcome but also sticking in the velocity regime studied was observed (Fig.\ref{collisions}). Detailed further studies of collisions are needed taking also into account repulsive and attractive forces during the actual collision to draw a conclusion of a threshold velocity between sticking and rebound. A related topic are slow collisions at high temperature as they might occur close to a star. The 
sticking properties of dust aggregates change, especially close to melting point. Since the levitation process 
is providing lift for all temperatures larger than the initial lifting temperature, collisional studies
can be carried out at relevant and varying high temperatures. We also envision explicit studies of the thermal, thermophoretic and photophoretic forces of dust aggregates 
which are strongest in the pressure regime (mbar) at which the levitation occurs. 
First tests show that levitated dust aggregates readily change their lateral motion and rotation
if illuminated externally which we attribute to photophoresis. 

\begin{acknowledgments}
We thank the referees for a constructive review. This work is funded by the Deutsche Forschungsgemeinschaft.
\end{acknowledgments}
\bibliography{kelling}
\end{document}